\documentclass[preprint,12pt,authoryear]{elsarticle}

\usepackage{epsfig}

\usepackage{amssymb}
\usepackage{amsmath} 
\usepackage{multirow}

\usepackage{longtable}

\usepackage{amsthm}

\journal{Experts Systems with Applications}

\begin{document}

\begin{frontmatter}



\begin{graphicalabstract}
\begin{figure}[htbp]
    \centering
    \includegraphics[width=1.0\textwidth]{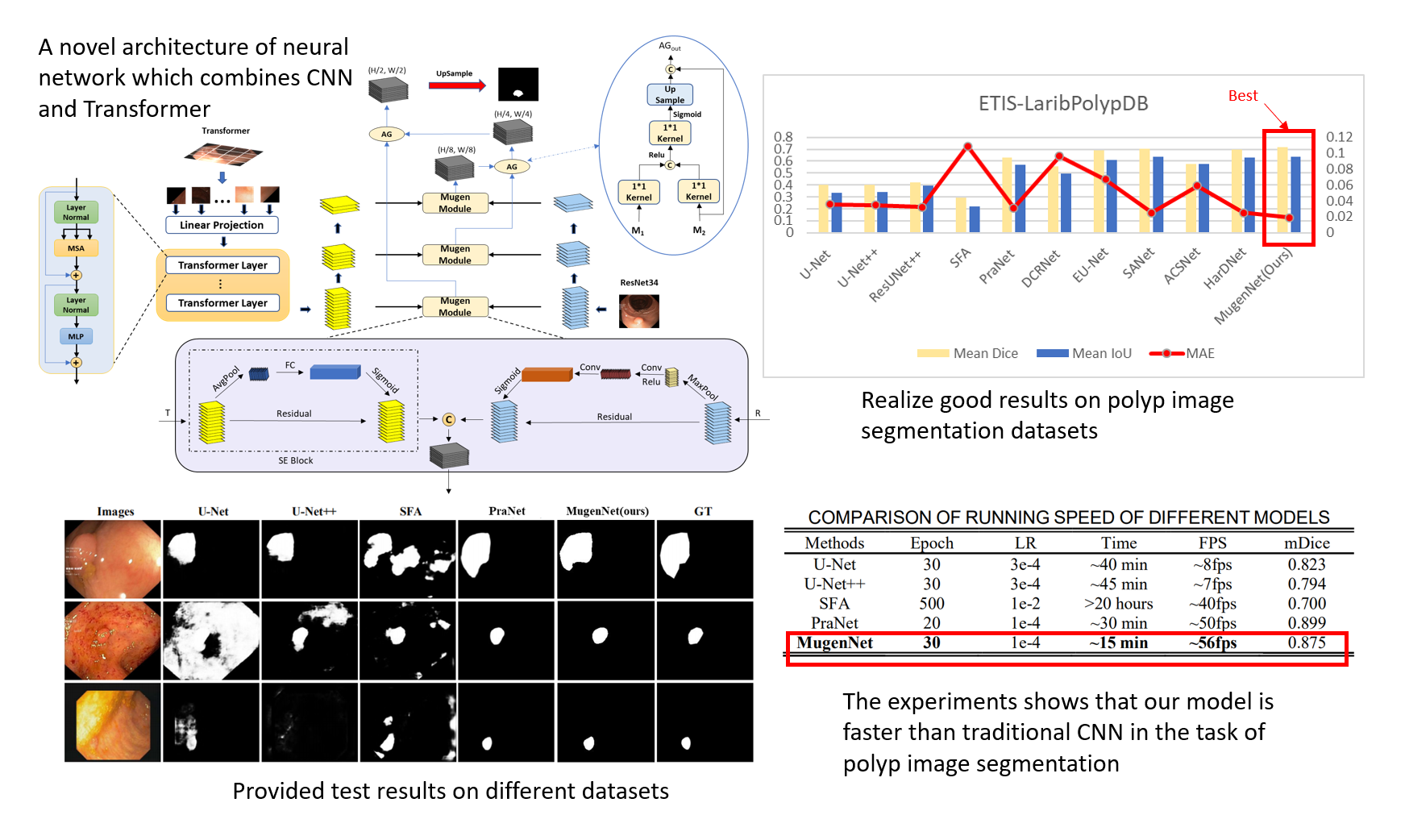}
    \label{fig fig.0}
\end{figure}
\end{graphicalabstract}

\begin{highlights}
\item Segmentation by combining convolutional neural network and transformer network.
\item Novel application of the novel segmentation method for colonic image processing.
\item Comprehensive justification and experimentation for the novel segmentation method.
\end{highlights}

\title{MugenNet: A Novel Combined Convolution Neural Network and Transformer Network with its Application for Colonic Polyp Image Segmentation}

 \author[label1]{Chen Peng}
 \ead{ceciliapengchen@163.com}
 \author{Zhiqin Qian\corref{cor1}\fnref{label1}}
 \ead{qianzhiqin@ecust.edu.cn}
 \cortext[cor1]{Zhiqin Qian, Wenjun Zhang}
 \author[label1]{Kunyu Wang}
 \ead{x849545512@hotmail.com}
 \author[label1]{Qi Luo}
 \ead{luoqi_ecust@163.com}
 \author[label2]{Zhuming Bi}
 \ead{biz@pfw.edu}
 \author{Wenjun Zhang\corref{cor1}\fnref{label3}}
 \ead{chris.zhang@usask.ca}
 \affiliation[label1]{organization={East China University of Science and Technology},
             postcode={200237},
             state={Shanghai},
             country={China}}

 \affiliation[label2]{organization={Purdue University},
             city={West Lafayette},
             postcode={47907},
             state={Indiana},
             country={America}}

 \affiliation[label3]{organization={University of Saskatchewan},
             city={Saskatoon},
             postcode={SK S7N 5A9},
             state={Saskatchewan},
             country={Canada}}             

%% \author{}

%% \affiliation{organization={},%Department and Organization

\begin{abstract}
Biomedical image segmentation is a very important part in disease diagnosis. The term "colonic polyps" refers to polypoid lesions that occur on the surface of the colonic mucosa within the intestinal lumen. In clinical practice, early detection of polyps is conducted through colonoscopy examinations and biomedical image processing. Therefore, the accurate polyp image segmentation is of great significance in colonoscopy examinations. Convolutional Neural Network (CNN) is a common automatic segmentation method, but its main disadvantage is the long training time. Transformer utilizes a self-attention mechanism, which essentially assigns different importance weights to each piece of information, thus achieving high computational efficiency during segmentation. However, a potential drawback is the risk of information loss. In the study reported in this paper, based on the well-known hybridization principle, we proposed a method to combine CNN and Transformer to retain the strengths of both, and we applied this method to build a system called MugenNet for colonic polyp image segmentation. We conducted a comprehensive experiment to compare MugenNet with other CNN models on five publicly available datasets. The ablation experiment on MugentNet was conducted as well. The experimental results show that MugenNet achieves significantly higher processing speed and accuracy compared with CNN alone. The generalized implication with our work is a method to optimally combine two complimentary methods of machine learning.
\end{abstract}


\end{frontmatter}


\section{Introduction}
Colonoscopy is an effective technique for detecting colorectal polyps, which are polypoid lesions on the surface of the colonic mucosa within the intestinal lumen. Colorectal polyps commonly lead to intestinal bleeding and inflammatory responses, and in severe cases, they may progress to colon cancers. Colonoscopy is the most effective way to prevent colon cancers, but the procedure may cause the risk of perforation and severe discomfort to the patient \citep{r7,r20}. The detection of abnormal polyps is by means of image processing in colonoscopy.

Previous research indicated that the clinical miss rate for colonic polyps can be as high as 25\% \citep{r13}. The segmentation of polyps from colonoscopy images is the first step in early detection of colon cancers.  Due to the variation in size, color, and texture among polyps of similar types, and the unclear boundaries between polyps and the surrounding mucosa, polyp segmentation remains a challenging task.

Convolutional neural network (CNN) is a well-known image segmentation method that extracts features from pixels or raw information blocks using convolutional operators \citep{r19}. However, the shortcoming with CNN includes (1) computational overhead in training of CNN, (2) not effectiveness in dealing with the biomedical images with low resolution, and (3) potential to be over-fitting \citep{r38}. 

Transformer is a machine learning process, which takes a series of symbols as the input, and produces the semantics from this series as the output \citep{r28}. Transformer is based on the self-attention mechanism. The self-attention mechanism refers to giving more attention to the relevance of a chunk of information. Suppose that there are three pieces of information, A, B, C. If A is closer to B than to C, a higher attention would be given to B with respect to A. 

To a 2D image, CNN does not have different attentions in the scanning process of the chunks of information \citep{r40}. This feature with CNN has two consequences as opposed to Tramsformer: (1) the computational cost with Transformer will be much lower than that with CNN \citep{r41}, because CNN must have a considerable amount of redundant computations, (2) Transformer is free of the overfitting problem, suggesting that it be more accurate than CNN. 

However, the shortcoming with the Transformer/self-attention mechanism is that the approach is highly dependent on the assessment of the relevance, which is further specific to different applications. For instance, in the system called ImageNet \citep{r8}, such an assessment must rely on external datasets, which is in essence a kind of pre-training process. This shortcoming with Transformer contrasts with the merit with CNN (i.e., the equity among all chunks of information). Therefore, it is a promising idea to combine CNN and Transformer, because the two are complementary, according to the so-called engineering hybridization principle \citep{r39}.

Currently, there is limited research on effectively combining CNN and Transformer models. Among these studies, \cite{r4} proposed TransUNet for organ segmentation in biomedical images. They accomplished the combination of CNN and Transformer by employing a chain-like structure, such as CNN-Transformer-CNN-Transformer and so forth. Their method is well-suited for organ recognition tasks because organ boundaries are relatively clear, and uncertainty levels are relatively low. The definition of uncertainty shall be referred to \cite{r3}. However, their system can’t perform parallel processing, which is one of the important merits of the Transformer as well. In another work, \cite{r41} proposed a combined system called Transfuse. Their way of combination is that first, CNN and Transformer process separately, and then they combine two feature maps, created by each of them, respectively. The results show that their model can partly overcome the shortcomings with CNN (i.e., gradient disappearance and feature reuse). In our observation, their approach may exhibit the overfitting problem when the number of epochs is high. 

The basic idea behind the work reported in this paper is to combine or fuse Transformer and CNN due to their nearly complementary ability according to the hybridization principle \citep{r39}. Specifically, we constructed a new neural network which combines Transformer and CNN for polyp segmentation, named MugenNet. We conducted comprehensive experiments with the MugenNet on five public colon polyp datasets using six performance indicators or indexes. We also compared MugenNet with other ten neural networks. These models are listed in Section 3. The result of the experiments shows that  (1) our model can achieve a mean Dice of 0.714 on the most challenging ETIS dataset, specifically 13.7\% improvement over the current state of the art CNN model (i.e., PraNet), and (2) our model can achieve a better accuracy and faster image processing speed than the traditional CNN, specifically the speed of our model being 12\% higher than that of PraNet (only taking 30 epochs within 15 minutes to complete the fitting of the model on the training dataset). 

\begin{figure}[htbp]
    \centering
    \includegraphics[width=1.0\textwidth]{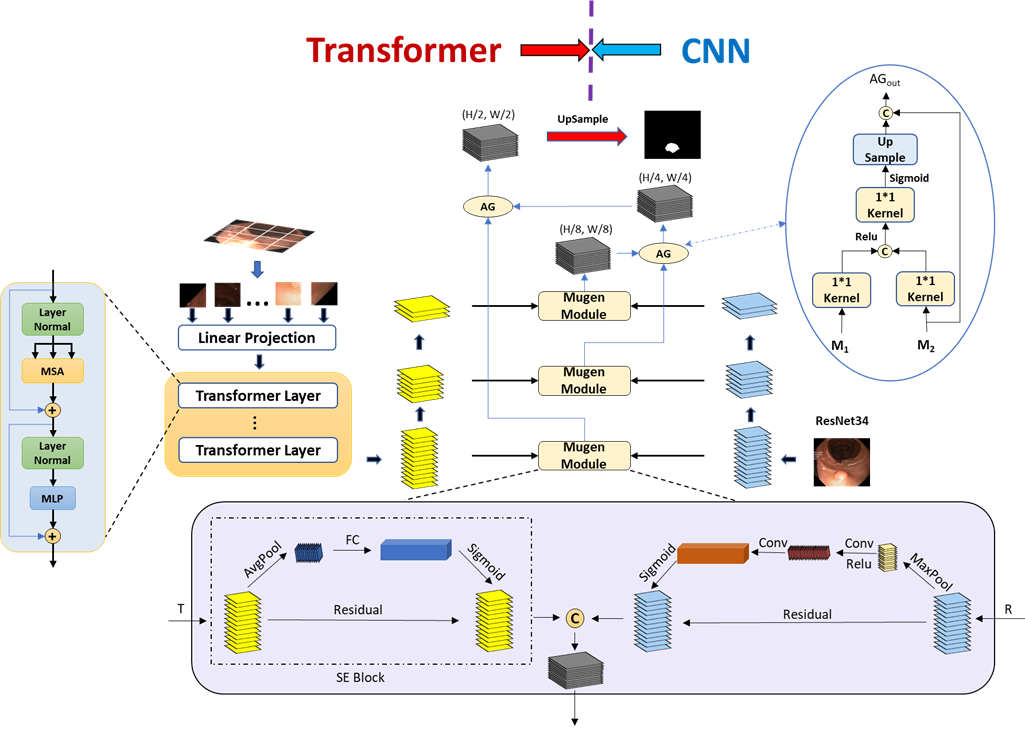}
    \caption{The architecture of MugenNet: combine Transformer branch (left) with CNN branch (right) in Mugen Module (middle). FC: Fully Connected. AG: Attention Gate. SE: Squeeze and Excitation.}
    \label{fig fig.1}
\end{figure}

Therefore, the main contribution of our work lies in the development of a novel idea and the corresponding method, which combines CNN and Transformer by adhering to the principles of hybrid engineering \citep{r39}. We successfully implemented this idea and method, resulting in a model called MugenNet for colonic polyp image segmentation. Furthermore, the ideas and methods proposed in this paper should be applicable to combinations of different machine learning algorithms and other image processing applications. 

The remaining part of the paper is organized as follows. Section 2 discusses the idea, method, and their implementation into the model MugenNet. Section 3 presents the experiment to validate MugenNet. Section 4 is the conclusion along with the discussion of some future works.

\section{The Proposed MugenNet System}
MugenNet is mainly composed of a CNN decoder based on Resnet-34 and a Transformer decoder based on ViT, two of which are run in parallel, as shown in Figure 1. The following sections will discuss MugenNet in detail.

\subsection{Design of the CNN branch}

ResNet-34 \citep{r14} was chosen as the CNN branch upon a trade-off between retaining more local information and losing more local information during a training process. We used two up-sampling operations to generate three matrices, which stored the feature maps of different scales; as depicted in the inset in the upper right part of Figure 1. These feature maps can be utilized to progressively restore the spatial resolution in subsequent processes. The matrix obtained through the convolution operation in the CNN branch (the right side of Figure 1) served as the input to the Mugen module and fused with the result obtained from the Transformer branch, which will be discussed in the next section. 

\subsection{Design of the Transformer Branch}

ViT was used as the Transformer branch in our model. The input image was divided into 16 blocks, and a self-attention operation was performed on each block \citep{r9}. We used two up-sampling operations to generate feature maps, matching their dimensions with the results obtained from the CNN branch. Assuming the input image is $x \in R^{H\times W\times C} $ where H and W represent the length and the width of the original image respectively, and C is the channel of the image (if it is a color image, $C= 3$, otherwise $c= 1$). First, divide $x$ into $4\times4$ patches, and every patch is $x_p\in R^{N\times(P^2\times C)}$, where P is the side length of $x_p$, $(P,P)$ is the resolution of each patch, and $N=HW/P^2$, indicating the number of patches. Then, add a location code to every patch and put each patch into a sequence as the Transformer modules (see the left part of Figure 1). Therefore, each Transformer module consisted of Layer Normalization (LN), Multi-head Self-Attention (MSA) and Multi-layer Perceptron (MLP), as depicted in the left part of Figure 1. The use of LN instead of Batch Normalization was because LN helped to increase the stability during training with a high learning rate \citep{r35,r41}. MSA was used to calculate the relevant information learned in different subspaces. MLP was a stack of activation functions and served as the output layer of CNN.

The activation function used is the so-called softmax, which was used to compute the three feature maps (Q, K and V), i.e., 
\begin{equation}
 N_{out}=softmax(\frac{q_ik^T}{\sqrt{D_h}})v   
\end{equation}
In Equation (1), $N_{out}$ represents the output of each attention module, which was then input to the Mugen module. The post-processing is referred to \cite{r21}. Specifically, two consecutive up-sampling convolution layers were connected to the output layer of the Transformer branch to recover the spatial resolution. The feature maps of different dimensions were preserved and fused with the output of the CNN branch. 

\subsection{Mugen Module}

The Mugen module was to combine the global chunks of information from Transformer and CNN, respectively. The process of the combination was composed of two parts. The first part was the Squeeze and Excitation (SE) block \citep{r21}, and the second part was the module that was built based on the channel attention. SE-Block can recalibrate the corresponding characteristics of the channels by explicitly modeling the inter-dependence between the channels and can improve the performance of the neural network without increase of the computational overhead. Channel Attention mechanism learns the weights of importance for each channel, allowing it to dynamically adjust the feature responses of each channel. This enables the network to effectively capture and utilize information from different channels. Such a mechanism aids in improving the performance of the network for specific tasks, and has shown significant improvements in various visual tasks including image classification, object detection, and semantic segmentation.

In the CNN branch, we adopted the channel attention module \citep{r32}. After global max-pooling, the feature map was input into the neural network consisting of two hidden layers. The number of neurons in the first layer was $C/r$ and the number of neurons in the second layer was $C$, where $C$ was the number of channels and $r$ was the reduction rate. We removed the  the global average-pooling part from the Convolutional Block Attention Module (CBAM) to focus more on the pixel-level edge information in the image segmentation task. Finally, we used the sigmoid function to generate the channel attention features.

After the above treatment, two feature maps were obtained: one from the Transformer branch and the other from the CNN branch, respectively. Inspired by ResNet, we added a residual connection architecture (Figure 1) before the final output from the Mugen module. 
ResNet helps to alleviate the vanishing or exploding gradient problem during the training process.

The steps in the Mugen module include the following computations:
\begin{equation}
\begin{cases}
&{\hat{t}}^i=SeAtten(t^i)\\
&{\hat{r}}^i=ChannelAtten(r^i)\\
&{z^i}_{out}=Residual(\left[{\hat{t}}^i,{\hat{r}}^i,t^i,r^i\right])\\
\end{cases}
\end{equation}

After the Mugen module, a stack of hidden layers with convolution and batch normalization was used to reduce the output dimension to one. The characteristic graph was used as the intermediate output ${z^i}_{out}$. We used the progressive up-sampling method to process the characteristic graph obtained from the Transformer and CNN branches. 

As shown in Figure 1, initially there were $t^0,\ r^0\in R^{\frac{H}{16}\ast\frac{W}{16}\ast D_0}$. After two consecutive up-sampling operations, we obtained $t^1,\ r^1\in R^{\frac{H}{8}\ast\frac{W}{8}\ast D_1}$ and $t^2,\ r^2\in R^{\frac{H}{4}\ast\frac{W}{4}\ast D_2}$, which restored the spatial resolution to the same dimension as the input image. Finally, we input $t^i,\ r^i$ into the three Mugen modules respectively, perform feature fusion, obtain three feature maps, forming the semantic segmentation map.

\subsection{Details of the transfer and loss function}
In the Mugen module, we adopted an architecture to combine Transformer and CNN, called Feature Pyramid (FP) along with the residual, proposed by \cite{r6}. After the top-down aggregation, the residual architecture was used to gather strong features, which was expected to improve the accuracy of the model. Compared with other bi-directional methods, the residual FP can more effectively achieve high-precision detection of targets.

From the Mugen module, we can obtain three feature maps of different dimensions. Then, Attention Gate (AG) was used to restore the resolution of attention. It is worth noting that, according to \cite{r23}, AG can focus on target architectures of different shapes and sizes, suppress irrelevant regions in the input image, and highlight the target features of specific tasks \citep{r31}. The architecture of AG is shown in Figure 1. 

\begin{figure}[htbp]
    \centering
    \includegraphics[width=1.0\textwidth]{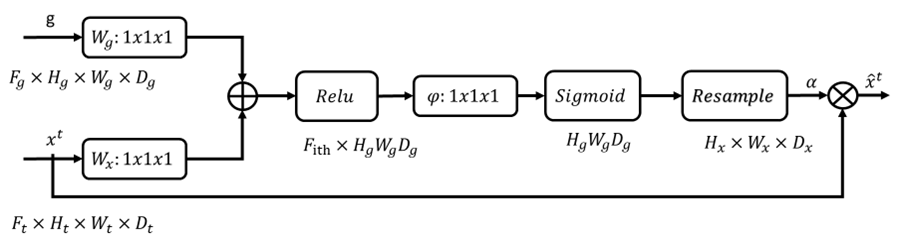}
    \caption{The architecture of attention gate.}
    \label{fig fig.2}
\end{figure}

We used two up-sampling operations to gradually restore the original resolution of the feature map. First, we set the output of the first Mugen module $({y^1}_{out})$ equal to ${z^1}_{out}\in R^{\frac{H}{8}\ast\frac{W}{8}\ast D_1}$. Then, we put ${z^1}_{out}$ and the output of the second Mugen module $({y^2}_{out})$ into the first attention gate to obtain ${z^2}_{out}\in R^{\frac{H}{4}\ast\frac{W}{4}\ast D_2}$. After that, we put ${z^2}_{out}$ and the output of the third Mugen module $({y^3}_{out})$ into the second attention gate to obtain ${z^3}_{out}\in R^{\frac{H}{2}\ast\frac{W}{2}\ast D_3}$. Finally, we conducted a separate up-sampling of ${z^3}_{out}$ to obtain the final semantic segmentation graph $z_{out}\in R^{H\ast W\ast D}$.

We defined the loss function as $L=L_{IoU}^\omega+n\ast L_{BCE}^\omega$ during the training process, according to \cite{r23,r26}, where $L_{IoU}^\omega$ was the global weighted average $IoU$ loss (added to the global constraint, and $L_{BCE}^\omega$ was the weighted binary cross entropy loss (added to the pixel level constraint). Further, $n$ was a hyperparameter used to allocate the proportion of the $IoU$ loss and the $BCE$ loss in the total loss $L$. Compared with the standard $IoU$ loss, which is commonly used in the semantic segmentation tasks, the global average $IoU$ loss highlights its the importance of complex sample pixels by increasing the their weight of the complex sample pixels. At the same time, the weighted binary cross entropy loss pays more attention to focused more on the complex sample pixels, which can effectively extract the pixel-level edge features at the pixel level. Considering that the goal of the model was of the colonic polyp segmentation and the edge features were not obvious prominent, we took set n to be $6/5$ to increase the proportion of the pixel level constraints in the overall-total loss function. $S_x^{up}$ was up-sampled to the same size as the ground truth graph image $G$. The proposed overall-loss function of MugenNet for colonic polyp detection can then be expressed by:
\begin{equation}
L_{total}=\alpha L\left(G,S_t^{up}\right)+\beta L\left(G,S_r^{up}\right)+\gamma L\left(G,S_z^{up}\right)    
\end{equation}
where $S_t^{up}$ and $S_r^{up}$ represent the characteristic graph obtained from the Transformer branch and CNN branch, respectively. $S_z^{up}$ represents the output of the semantic segmentation graph after up-sampling. The parameters $\alpha, \beta, \gamma$ are adjustable hyperparameters.

\subsection{Implementation}
We implemented the MugenNet module (referred to as the model hereafter) in the PyTorch framework on an NVIDIA RTX 3080 GPU. The resolution of input images was shaped to $256\times192$. The dataset was divided into training, validation, and test data sets in the ratio of $7:2:1$. We used the Adam optimizer to adjust the overall parameters, with a learning rate of $1\times{10}^{-4}$. The batch size was set to 16. The final prediction result was generated after the sigmoid operation.

\section{Experiment}
We compared the MugenNet model with some existing CNN models, which belong to the category of CNN models in terms of the abilities such as learning, generalization, and qualitative segmentation.

\subsection{The dataset and model}
The experiment was conducted on five publicly available polyp segmentation datasets, namely CVC-300 \citep{r29}, CVC-ClinicDB \citep{r1}, CVC-ColonDB \citep{r26}, ETIS \citep{r24}, Kvasir \citep{r16}. We randomly selected 1450 images from Kvasir and CVC-ClinicDB to train our model, with the weights initialized based on the training weights of DeiT-small \citep{r27} and resnet-34 \citep{r14}. We then compared our model with ten other convolutional neural network models for biomedical image segmentation, namely U-Net \citep{r22}, U-Net++ \citep{r44}, ResUNet++ \citep{r17}, SFA \citep{r12}, PraNet \citep{r11}, DCRNet \citep{r37}, EU-Net \citep{r19}, SANet \citep{r30}, ACSNet \citep{r38}, and HaeDNet \citep{r15}. 

\subsection{Training setting and performance indicators}
To assess the generalization ability of the model, besides the two datasets used for training the model, we also tested it on three additional datasets, namely CVC-300, CVC-ColonDB and ETIS. 

\subsection{Performance indicators}
To evaluate the proposed MugenNet for the semantic segmentation of colon polyp images, we used the performance indicators of $mIoU$, $mDice$, $MAE$ loss, $F_\beta^\omega$, $S_\alpha$ and $E_\xi$. The mathematical expression for the definition of each performance indicator along with its significance in the image segmentation task is introduced below.

The full name of $IoU$ is Intersection over Union, which represents the ratio of intersection and union between the bounding box and ground truth. The average $IoU$ ($mIoU$) can be calculated by
\begin{equation}
mIoU=\frac{1}{n}\sum_{i=0}^{n}\frac{A_i\cap B_i}{A_i\cup B_i}
\end{equation}

During the training process of a semantic segmentation model, the value of the $IoU$ loss will be used to evaluate whether the prediction result of the model is good or valid. If $IoU>0.5$, the prediction is considered valid. Additionally, we also calculated the $IoU$ value relative to the ground truth for each training batch, and we then took the average as the $mIoU$ loss for this specific training batch.

The Dice coefficient is a measure of similarity between two sets (say A and B) and is used for medical image segmentation. The formula for the mean Dice coefficient ($mDice$) is as follows:
\begin{equation}
mDice=\frac{1}{n}\sum_{i=0}^{n}\frac{2\left|A_i\cap B_i\right|}{\left|A_i\right|+\left|B_i\right|}
\end{equation}

The range of the Dice coefficient is (0, 1). The closer the coefficient is to 1, the higher the similarity between sets A and B. We calculated the Dice coefficient between the model’s predicted results and the ground truth to evaluate the reliability of the model in colon polyp image segmentation.

\begin{figure}[htbp]
    \centering
    \includegraphics[width=1.0\textwidth]{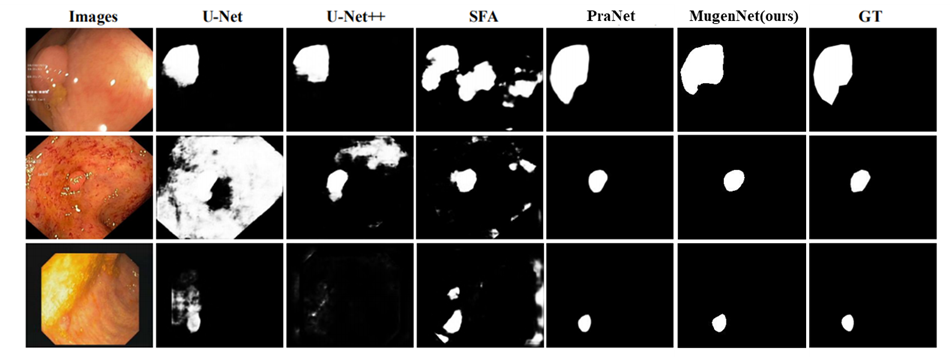}
    \caption{Comparison of the MugenNet with the other four nets (U-Net, U-net++, SFA, Pranet) on the Kvasir dataset.}
    \label{fig fig.3}
\end{figure}

The full name of $MAE$ is the Mean Absolute Error, which is used to calculate the error between predicted values and the ground truth. $MAE$ is calculated by
\begin{equation}
MAE=\frac{1}{n}\sum_{i=0}^{n}\left|A_i-B_i\right|
\end{equation}
The advantage of $MAE$ is that it is insensitive to outliers. In this paper, we took it as one of the criteria to evaluate the reliability of prediction results.

The parameters for precision and recall metrics include: TP (True Positive), TN (True Negative), FP (False Positive), and FN (False Negative). We compared each pixel in the map of the prediction result with the ground truth and classified the prediction result into (i) correct $(D\left(i\right)=G(i))$ or (ii) incorrect $(D\left(i\right)\neq G(i))$. The four attributes are then calculated by
\begin{equation}
\begin{cases}
&TP=D\ast G\\
&TN=(1-D)\ast(1-G)\\
&FP=D\ast(1-G)\\
&FN=(1-D)\ast G\\
\end{cases}
\end{equation}
Then, we calculated the precision and recall by
\begin{equation}
\begin{cases}
&Precision=\ \frac{TP}{TP+FP}\\
&Recall=\ \frac{TP}{TP+FN}\\
\end{cases}
\end{equation}

We also used the weighted F-measure $(F_\beta^\omega)$ \citep{r18}, S-measure $(S_\alpha)$ \citep{r5} and E-measure $(E_\xi)$ \citep{r10} to evaluate the model performances, similar to \cite{r11}. The choice of these three metrics is to evaluate the precision and recall of the model during testing, as well as to assess the model's performance. The formulas for these three metrics are as follows.

We can calculate the weighted F-measure by
\begin{equation}
F_\beta^\omega=\frac{\left(1+\beta^2\right){Precision}^\omega\cdot{Recall}^\omega}{\beta^2\cdot{Precision}^\omega+{Recall}^\omega}
\end{equation}
where $\omega$ represents weight, $\beta$ is the adjustable coefficient. In this study, we took $\omega=1,\ \beta=1/2$. It is worth noting that the weighted F-measure is a robust metric because it incorporates both precision and recall. The S-measure is calculated by
\begin{equation}
\begin{cases}
&S=\alpha\cdot S_0+(1-\alpha)\cdot S_r\\
&S_0=\mu\cdot S_{FG}+(1-\mu)\cdot S_{BG}\\
&S_{FG}=\frac{2{\bar{x}}_{FG}}{{({\bar{x}}_{FG})}^2+1+2\lambda\cdot\sigma_{x_{FG}}}\\
&S_{BG}=\frac{2{\bar{x}}_{BG}}{{({\bar{x}}_{BG})}^2+1+2\lambda\cdot\sigma_{x_{BG}}}\\
\end{cases}
\end{equation}
where FG stands for Foreground and BG stands for Background. We set $\alpha=0.5,\mu=0.5,\lambda=1$. $x_{FG}$ represents the probability of the foreground region in the predicted result and the ground truth. $x_{BG}$ represents the probability of the background region in the predicted result and the ground truth. S-measure will be used to assess the structural similarity of targets in the region. 

\begin{figure}[htbp]
    \centering
    \includegraphics[width=1.0\textwidth]{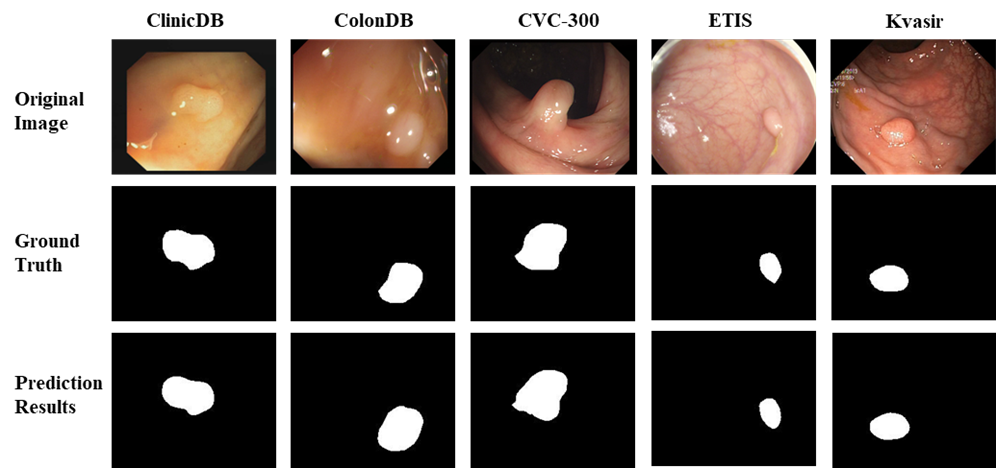}
    \caption{Comparison of the performance of MugenNet on the five datasets (ClinicDB, ColonDB, CVC 300, ETIS, Kvasir).}
    \label{fig fig.4}
\end{figure}

E-measure is calculated by
\begin{equation}
\begin{cases}
&E_\xi=\frac{1}{w\cdot h}\sum_{x=1}^{w}\sum_{y=1}^{h}\frac{2\varphi_{GT}\cdot\varphi_{FM}}{\varphi_{GT}\cdot\varphi_{GT}+\varphi_{FM}\cdot\varphi_{FM}}\\
&\varphi_I=I-\mu_I\cdot A\\
\end{cases}
\end{equation}
where $I$ is the input binary foreground map. $\mu_I$ is the global mean of $I$. $A$ is a matrix where all elements have a value of 1, and its size is to the same as that of matric $I$. The E-measure was used to evaluate the performance of the model at the pixel level. We took the mean value of E-measure. 

\subsection{Comparison of the performance}
The results of the comparison are shown in Table \uppercase\expandafter{\romannumeral 1}, where "Null" means that the corresponding results are not provided in the literature. The data of other existing models were obtained from the corresponding literature and their implementation was done using their published code. 

\begin{table}[htbp]
\centering
\caption{COMPARISON RESULTS OF CVC-300}
\begin{tabular}{ccccccc}
\hline
\multirow{2}{*}{Methods} & \multicolumn{6}{c}{CVC-300 dataset}                                                                                                                                                                                                                \\ \cline{2-7} 
                         & \multicolumn{1}{c}{mDice} & \multicolumn{1}{c}{mIoU}  & \multicolumn{1}{c}{MAE}   & \multicolumn{1}{c}{$F_\beta^\omega$} & \multicolumn{1}{c}{$S_\alpha$} & {$E_\xi$} \\ \hline
U-Net                    & \multicolumn{1}{l}{0.710}          & \multicolumn{1}{l}{0.627}          & \multicolumn{1}{l}{0.022}          & \multicolumn{1}{l}{0.567}                                                          & \multicolumn{1}{l}{0.793}                    & 0.826                                     \\
U-Net++                  & \multicolumn{1}{l}{0.707}          & \multicolumn{1}{l}{0.624}          & \multicolumn{1}{l}{0.018}          & \multicolumn{1}{l}{0.581}                                                          & \multicolumn{1}{l}{0.796}                    & 0.831                                     \\
ResUNet++                & \multicolumn{1}{l}{0.763}          & \multicolumn{1}{l}{0.701}          & \multicolumn{1}{l}{0.021}          & \multicolumn{1}{l}{0.613}                                                          & \multicolumn{1}{l}{0.801}                    & 0.867                                     \\
SFA                      & \multicolumn{1}{l}{0.467}          & \multicolumn{1}{l}{0.329}          & \multicolumn{1}{l}{0.065}          & \multicolumn{1}{l}{0.341}                                                          & \multicolumn{1}{l}{0.640}                    & 0.644                                     \\
PraNet                   & \multicolumn{1}{l}{0.871}          & \multicolumn{1}{l}{0.797}          & \multicolumn{1}{l}{0.010}          & \multicolumn{1}{l}{0.843}                                                          & \multicolumn{1}{l}{0.925}                    & 0.950                                     \\
DCRNet                   & \multicolumn{1}{l}{0.856}          & \multicolumn{1}{l}{0.788}          & \multicolumn{1}{l}{0.016}          & \multicolumn{1}{l}{0.830}                                                          & \multicolumn{1}{l}{0.921}                    & 0.943                                     \\
EU-Net                   & \multicolumn{1}{l}{0.837}          & \multicolumn{1}{l}{0.765}          & \multicolumn{1}{l}{0.015}          & \multicolumn{1}{l}{0.805}                                                          & \multicolumn{1}{l}{0.904}                    & 0.919                                     \\
SANet                    & \multicolumn{1}{l}{\textbf{0.888}} & \multicolumn{1}{l}{\textbf{0.815}} & \multicolumn{1}{l}{0.018}          & \multicolumn{1}{l}{0.859}                                                          & \multicolumn{1}{l}{\textbf{0.928}}           & 0.962                                     \\
ACSNet                   & \multicolumn{1}{l}{0.732}          & \multicolumn{1}{l}{0.627}          & \multicolumn{1}{l}{0.016}          & \multicolumn{1}{l}{0.703}                                                          & \multicolumn{1}{l}{0.837}                    & 0.871                                     \\
HarDNet                  & \multicolumn{1}{l}{0.874}          & \multicolumn{1}{l}{0.804}          & \multicolumn{1}{l}{0.019}          & \multicolumn{1}{l}{0.852}                                                          & \multicolumn{1}{l}{0.924}                    & 0.948                                     \\ 
\textbf{MugenNet(Ours)}           & \multicolumn{1}{l}{0.863}          & \multicolumn{1}{l}{0.795}          & \multicolumn{1}{l}{\textbf{0.010}} & \multicolumn{1}{l}{\textbf{0.864}}                                                 & \multicolumn{1}{l}{0.921}                    & \textbf{0.972}
\\ \hline
\end{tabular}
\end{table}

From the results on the five different datasets (Table \uppercase\expandafter{\romannumeral 1} to Table \uppercase\expandafter{\romannumeral 5}), it can be found that the proposed MugenNet shows execellent performance in the CVC-ColonDB and ETIS polyp segmentation datasets (Table \uppercase\expandafter{\romannumeral 2} and Table \uppercase\expandafter{\romannumeral 3}), showing the best overall performance across the three metrics. It is worth mentioning that our training datasets are sourced from Kvasir and CVC-ClinicDB (Table \uppercase\expandafter{\romannumeral 4} and Table \uppercase\expandafter{\romannumeral 5}), indicating that our model has good generalization ability and is suitable for the predicting semantic segmentation results on unknown datasets. Although the results on the other three datasets indicate that our model has not achieved the best performance but not far from PraNet (only short of 1.6\%), which is the current SOTA (the state of the art) performance. We compared the inference time of the different models in Table \uppercase\expandafter{\romannumeral 6}. The results show that with the same training dataset, our model has reduced the inference time by 12\% compared with PraNet. At the same time, our model significantly outperforms traditional biomedical image semantic segmentation model (such as U-Net); especially on the ETIS dataset, our model's performance improvement reaches as high as 13.7\% compared with the PraNet model.

\begin{table}[ht]
\centering
\caption{COMPARISON RESULTS OF CVC-COLONDB}
\begin{tabular}{cllllll}
\hline
\multirow{2}{*}{Methods} & \multicolumn{6}{c}{CVC-ColonDB dataset}                                                                                                                                                                                                                                \\ \cline{2-7} 
                         & \multicolumn{1}{c}{mDice} & \multicolumn{1}{c}{mIoU}  & \multicolumn{1}{c}{MAE}   & \multicolumn{1}{c}{$F_\beta^\omega$} & \multicolumn{1}{c}{$S_\alpha$} & \multicolumn{1}{c}{$E_\xi$} \\ \hline
U-Net                    & \multicolumn{1}{l}{0.512}          & \multicolumn{1}{l}{0.444}          & \multicolumn{1}{l}{0.061}          & \multicolumn{1}{l}{0.498}                                                          & \multicolumn{1}{l}{0.712}                    & 0.696                                     \\
U-Net++                  & \multicolumn{1}{l}{0.483}          & \multicolumn{1}{l}{0.410}          & \multicolumn{1}{l}{0.064}          & \multicolumn{1}{l}{0.467}                                                          & \multicolumn{1}{l}{0.691}                    & 0.680                                     \\
ResUNet++                & \multicolumn{1}{l}{0.572}          & \multicolumn{1}{l}{0.501}          & \multicolumn{1}{l}{0.058}          & \multicolumn{1}{l}{0.576}                                                          & \multicolumn{1}{l}{0.706}                    & 0.701                                     \\
SFA                      & \multicolumn{1}{l}{0.469}          & \multicolumn{1}{l}{0.347}          & \multicolumn{1}{l}{0.096}          & \multicolumn{1}{l}{0.379}                                                          & \multicolumn{1}{l}{0.634}                    & 0.675                                     \\
PraNet                   & \multicolumn{1}{l}{0.712}          & \multicolumn{1}{l}{0.640}          & \multicolumn{1}{l}{0.043}          & \multicolumn{1}{l}{0.699}                                                          & \multicolumn{1}{l}{0.820}                    & 0.847                                     \\
DCRNet                   & \multicolumn{1}{l}{0.704}          & \multicolumn{1}{l}{0.631}          & \multicolumn{1}{l}{0.052}          & \multicolumn{1}{l}{0.684}                                                          & \multicolumn{1}{l}{0.821}                    & 0.840                                     \\
EU-Net                   & \multicolumn{1}{l}{0.756}          & \multicolumn{1}{l}{\textbf{0.681}} & \multicolumn{1}{l}{0.045}          & \multicolumn{1}{l}{0.730}                                                          & \multicolumn{1}{l}{0.831}                    & 0.863                                     \\
SANet                    & \multicolumn{1}{l}{0.753}          & \multicolumn{1}{l}{0.670}          & \multicolumn{1}{l}{0.043}          & \multicolumn{1}{l}{0.726}                                                          & \multicolumn{1}{l}{0.837}                    & 0.869                                     \\
ACSNet                   & \multicolumn{1}{l}{0.716}          & \multicolumn{1}{l}{0.649}          & \multicolumn{1}{l}{0.039}          & \multicolumn{1}{l}{0.697}                                                          & \multicolumn{1}{l}{0.829}                    & 0.839                                     \\
HarDNet                  & \multicolumn{1}{l}{0.735}          & \multicolumn{1}{l}{0.666}          & \multicolumn{1}{l}{0.038}          & \multicolumn{1}{l}{0.724}                                                          & \multicolumn{1}{l}{0.834}                    & 0.834                                     \\ 
\textbf{MugenNet(Ours)}           & \multicolumn{1}{l}{\textbf{0.758}} & \multicolumn{1}{l}{0.678}          & \multicolumn{1}{l}{\textbf{0.034}} & \multicolumn{1}{l}{\textbf{0.809}}                                                 & \multicolumn{1}{l}{\textbf{0.918}}           & \textbf{0.875}   
\\ \hline
\end{tabular}
\end{table}

The poor performance of models such as SFA and U-Net on the ETIS dataset suggests that they have weak generalization ability on unknown datasets. In contrast, our model outperforms traditional biomedical image segmentation models.

Figure 3 displays the polyp segmentation results of MugenNet on Kvasir dataset, comparing our model with U-Net, U-Net++, SFA and PraNet, where GT represents the ground truth. Our model can accurately locate polyps of different sizes and textures, then generate semantic segmentation maps. It is worth noting that the results of other models are compared with the results from \cite{r11}.

\begin{table}[htbp]
\centering
\caption{COMPARISON RESULTS OF ETIS}
\begin{tabular}{cllllll}
\hline
\multirow{2}{*}{Methods} & \multicolumn{6}{c}{ETIS dataset}                                                                                                                                                                                                                                \\ \cline{2-7} 
                         & \multicolumn{1}{c}{mDice} & \multicolumn{1}{c}{mIoU}  & \multicolumn{1}{c}{MAE}   & \multicolumn{1}{c}{$F_\beta^\omega$} & \multicolumn{1}{c}{$S_\alpha$} & \multicolumn{1}{c}{$E_\xi$} \\ \hline
U-Net                    & \multicolumn{1}{l}{0.398}          & \multicolumn{1}{l}{0.335}          & \multicolumn{1}{l}{0.036}          & \multicolumn{1}{l}{0.366}                                                          & \multicolumn{1}{l}{0.684}                    & 0.643                                     \\
U-Net++                  & \multicolumn{1}{l}{0.401}          & \multicolumn{1}{l}{0.344}          & \multicolumn{1}{l}{0.035}          & \multicolumn{1}{l}{0.683}                                                          & \multicolumn{1}{l}{0.683}                    & 0.629                                     \\
ResUNet++                & \multicolumn{1}{l}{0.423}          & \multicolumn{1}{l}{0.475}          & \multicolumn{1}{l}{0.029}          & \multicolumn{1}{l}{0.628}                                                          & \multicolumn{1}{l}{0.716}                    & 0.693                                     \\
SFA                      & \multicolumn{1}{l}{0.297}          & \multicolumn{1}{l}{0.217}          & \multicolumn{1}{l}{0.109}          & \multicolumn{1}{l}{0.231}                                                          & \multicolumn{1}{l}{0.557}                    & 0.531                                     \\
PraNet                   & \multicolumn{1}{l}{0.628}          & \multicolumn{1}{l}{0.567}          & \multicolumn{1}{l}{0.031}          & \multicolumn{1}{l}{0.600}                                                          & \multicolumn{1}{l}{0.794}                    & 0.808                                     \\
DCRNet                   & \multicolumn{1}{l}{0.556}          & \multicolumn{1}{l}{0.496}          & \multicolumn{1}{l}{0.096}          & \multicolumn{1}{l}{0.506}                                                          & \multicolumn{1}{l}{0.736}                    & 0.742                                     \\
EU-Net                   & \multicolumn{1}{l}{0.687}          & \multicolumn{1}{l}{0.609}          & \multicolumn{1}{l}{0.067}          & \multicolumn{1}{l}{0.636}                                                          & \multicolumn{1}{l}{0.793}                    & 0.807                                     \\
SANet                    & \multicolumn{1}{l}{0.705}          & \multicolumn{1}{l}{0.634}          & \multicolumn{1}{l}{0.025}          & \multicolumn{1}{l}{\textbf{0.685}}                                                 & \multicolumn{1}{l}{\textbf{0.849}}           & \textbf{0.881}                            \\
ACSNet                   & \multicolumn{1}{l}{0.578}          & \multicolumn{1}{l}{0.578}          & \multicolumn{1}{l}{0.059}          & \multicolumn{1}{l}{0.530}                                                          & \multicolumn{1}{l}{0.754}                    & 0.737                                     \\
HarDNet                  & \multicolumn{1}{l}{0.700}          & \multicolumn{1}{l}{0.630}          & \multicolumn{1}{l}{0.025}          & \multicolumn{1}{l}{0.671}                                                          & \multicolumn{1}{l}{0.828}                    & 0.854                                     \\ 
\textbf{MugenNet(Ours)}  & \multicolumn{1}{l}{\textbf{0.714}} & \multicolumn{1}{l}{\textbf{0.636}} & \multicolumn{1}{l}{\textbf{0.019}} & \multicolumn{1}{l}{0.606}                                                          & \multicolumn{1}{l}{0.677}                    & 0.831  
\\ \hline
\end{tabular}
\end{table}

\begin{table}[htbp]
\centering
\caption{COMPARISON RESULTS OF KVASIR}
\begin{tabular}{cllllll}
\hline
\multirow{2}{*}{Methods} & \multicolumn{6}{c}{KVASIR dataset}                                                                                                                                                                                                                                \\ \cline{2-7} 
                         & \multicolumn{1}{c}{mDice} & \multicolumn{1}{c}{mIoU}  & \multicolumn{1}{c}{MAE}   & \multicolumn{1}{c}{$F_\beta^\omega$} & \multicolumn{1}{c}{$S_\alpha$} & \multicolumn{1}{c}{$E_\xi$} \\ \hline
U-Net                    & \multicolumn{1}{l}{0.818}          & \multicolumn{1}{l}{0.746}          & \multicolumn{1}{l}{0.055}          & \multicolumn{1}{l}{0.794}                                                          & \multicolumn{1}{l}{0.858}                    & 0.881                                     \\
U-Net++                  & \multicolumn{1}{l}{0.821}          & \multicolumn{1}{l}{0.743}          & \multicolumn{1}{l}{0.048}          & \multicolumn{1}{l}{0.808}                                                          & \multicolumn{1}{l}{0.862}                    & 0.886                                     \\
ResUNet++                & \multicolumn{1}{l}{0.813}          & \multicolumn{1}{l}{0.793}          & \multicolumn{1}{l}{0.053}          & \multicolumn{1}{l}{0.831}                                                          & \multicolumn{1}{l}{0.873}                    & 0.878                                     \\
SFA                      & \multicolumn{1}{l}{0.723}          & \multicolumn{1}{l}{0.611}          & \multicolumn{1}{l}{0.075}          & \multicolumn{1}{l}{0.670}                                                          & \multicolumn{1}{l}{0.782}                    & 0.834                                     \\
PraNet                   & \multicolumn{1}{l}{0.898}          & \multicolumn{1}{l}{0.840}          & \multicolumn{1}{l}{0.030}          & \multicolumn{1}{l}{0.885}                                                          & \multicolumn{1}{l}{0.915}                    & 0.944                                     \\
DCRNet                   & \multicolumn{1}{l}{0.886}          & \multicolumn{1}{l}{0.825}          & \multicolumn{1}{l}{0.035}          & \multicolumn{1}{l}{0.868}                                                          & \multicolumn{1}{l}{0.911}                    & 0.933                                     \\
EU-Net                   & \multicolumn{1}{l}{\textbf{0.908}} & \multicolumn{1}{l}{\textbf{0.854}} & \multicolumn{1}{l}{0.038}          & \multicolumn{1}{l}{0.893}                                                          & \multicolumn{1}{l}{0.917}                    & 0.951                                     \\
SANet                    & \multicolumn{1}{l}{0.904}          & \multicolumn{1}{l}{0.847}          & \multicolumn{1}{l}{0.036}          & \multicolumn{1}{l}{0.892}                                                          & \multicolumn{1}{l}{0.915}                    & 0.949                                     \\
ACSNet                   & \multicolumn{1}{l}{0.898}          & \multicolumn{1}{l}{0.838}          & \multicolumn{1}{l}{0.032}          & \multicolumn{1}{l}{0.882}                                                          & \multicolumn{1}{l}{0.920}                    & 0.941                                     \\
HarDNet                  & \multicolumn{1}{l}{0.897}          & \multicolumn{1}{l}{0.839}          & \multicolumn{1}{l}{0.038}          & \multicolumn{1}{l}{0.885}                                                          & \multicolumn{1}{l}{0.912}                    & 0.942                                     \\ 
\textbf{MugenNet(Ours)}  & \multicolumn{1}{l}{0.888}          & \multicolumn{1}{l}{0.828}          & \multicolumn{1}{l}{\textbf{0.030}} & \multicolumn{1}{l}{\textbf{0.928}}                                                 & \multicolumn{1}{l}{\textbf{0.939}}           & \textbf{0.947} 
\\ \hline
\end{tabular}
\end{table}

\begin{table}[ht]
\centering
\caption{COMPARISON RESULTS OF CVC-CLINICDB}
\begin{tabular}{cllllll}
\hline
\multirow{2}{*}{Methods} & \multicolumn{6}{c}{CVC-CLINICDB dataset}                                                                                                                                                                                                                                \\ \cline{2-7} 
                         & \multicolumn{1}{c}{mDice} & \multicolumn{1}{c}{mIoU}  & \multicolumn{1}{c}{MAE}   & \multicolumn{1}{c}{$F_\beta^\omega$} & \multicolumn{1}{c}{$S_\alpha$} & \multicolumn{1}{c}{$E_\xi$} \\ \hline
U-Net                    & 0.823                     & 0.755                    & 0.019                   & 0.811                                                                              & 0.889                                        & 0.913                                     \\
U-Net++                  & 0.794                     & 0.729                    & 0.022                   & 0.785                                                                              & 0.873                                        & 0.891                                     \\
ResUNet++                & 0.796                     & 0.732                    & 0.036                   & 0.803                                                                              & 0.860                                        & 0.878                                     \\
SFA                      & 0.700                     & 0.607                    & 0.042                   & 0.647                                                                              & 0.793                                        & 0.840                                     \\
PraNet                   & 0.899                     & 0.849                    & \textbf{0.009}                   & 0.896                                                                              & 0.936                                        & 0.963                                     \\
DCRNet                   & 0.896                     & 0.844                    & 0.010                   & 0.890                                                                              & 0.933                                        & 0.964                                     \\
EU-Net                   & 0.892                     & 0.846                    & 0.011                   & 0.891                                                                              & 0.936                                        & 0.959                                     \\
SANet                    & \textbf{0.916}                     & 0.859                    & 0.012                   & 0.909                                                                              & 0.939                                        & \textbf{0.971}                                     \\
ACSNet                   & 0.882                     & 0.826                    & 0.011                   & 0.873                                                                              & 0.927                                        & 0.947                                     \\
HarDNet                  & 0.909                     & \textbf{0.864}                    & 0.018                   & 0.907                                                                              & 0.938                                        & 0.961                                     \\ 
\textbf{MugenNet(Ours)}  & 0.884                     & 0.823                    & 0.017                   & \textbf{0.947}                                                                              & \textbf{0.946}                                        & 0.969                                      \\ \hline                             
\end{tabular}
\end{table}

In addition to the comparison with the four existing models U-Net, U-Net++, SFA, PraNet in Figure 3 on the Kvasir dataset, we also tested our model on other four datasets (CVC-300, CVC-ClinicDB, CVC-ColonDB, ETIS-LaribPolypDB,). The results are shown in Figure 4. From Figure 4, our model (MugenNet) can accurately locate the position and size of polyps on the five different datasets. The results show that our model possesses stable and good generalization ability. 

We also trained our model (MugenNet) on the datasets of CVC-300, CVC-ClinicDB, CVC-ColonDB, ETIS-LaribPolypDB, with the results shown in Figure 5. It can be seen from Figure 5 that our model performs well when trained on the other four polyp datasets. This shows that our model is of good robustness (because its performance is insensitive to the training datasets) and can quickly adapt to multiple different datasets.

\begin{figure}[htbp]
    \centering
    \includegraphics[width=0.8\textwidth]{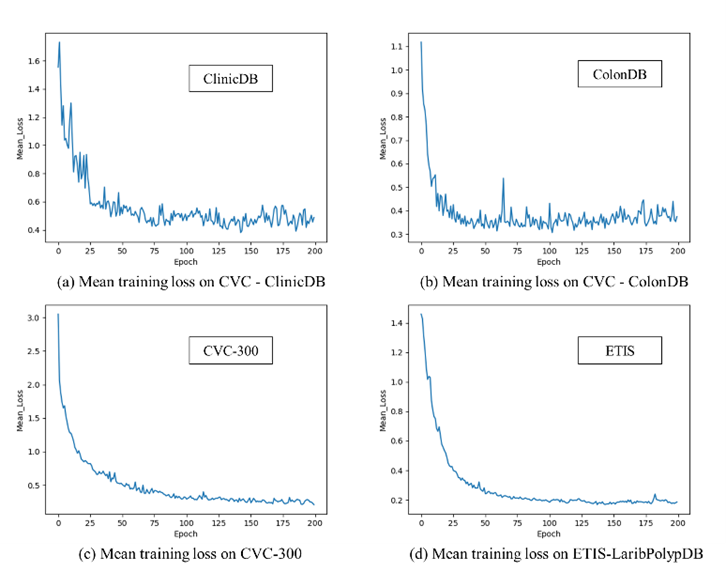}
    \caption{Training process on four datasets.}
    \label{fig fig.5}
\end{figure}

From the pre-training weight of the DeiT distillation neural network, as shown in Table \uppercase\expandafter{\romannumeral 6}, our model can converge at less than 30 epochs, and the training time is only about 15 minutes when the batch size is set to 16. $LR$ represents Learning Rate, $FPS$ represents Frames Per Second. Our model’s real-time running speed is approximately 56 frames per second (fps), which means that our model can be used to process video streaming data during colonoscopy examinations to perform real-time polyp image segmentation tasks.

We tested the Frames Per Second ($FPS$) of ten models on the training dataset to evaluate the inference speed of the models, as shown in Fig. 6. Since the FPS of U-Net and U-Net++ are both below 10 (as shown in Table VI), their inference speed is too slow, hence they are not shown in Fig. 6. It can be observed from Fig. 6 that the average $FPS$ of our model reaches 56. It is noted that when $FPS$ is greater than 30, the model was considered as reasonable for video stream data \citep{r11}. It can be seen from Fig. 6 and Table \uppercase\expandafter{\romannumeral 6} that our model (MugenNet) has advantages in processing video stream data for the colonic polyp image segmentation. 

The comparative results indicate that our model performs well on three unseen datasets, far surpassing traditional CNN-based biomedical image segmentation models such as U-Net, U-Net++, SFA and ResUNet++, and even slightly outperforming PraNet on certain datasets. 

It is worth mentioning that the models we compared are all based on convolutional neural networks, while our model (MugenNet) combines Transformer and CNN for colon polyp image segmentation. The experiment results show that the incorporation of Transformer can effectively improve the performance of neural networks, achieving faster processing speed. In essence, our model (MugenNet) possesses both the global learning capability of Transformer for capturing overall information and the local learning capability of CNN for focusing on specific regions. Therefore, our model achieves an excellent performance in the colon polyp image segmentation. 

\begin{figure}[ht]
    \centering
    \includegraphics[width=0.8\textwidth]{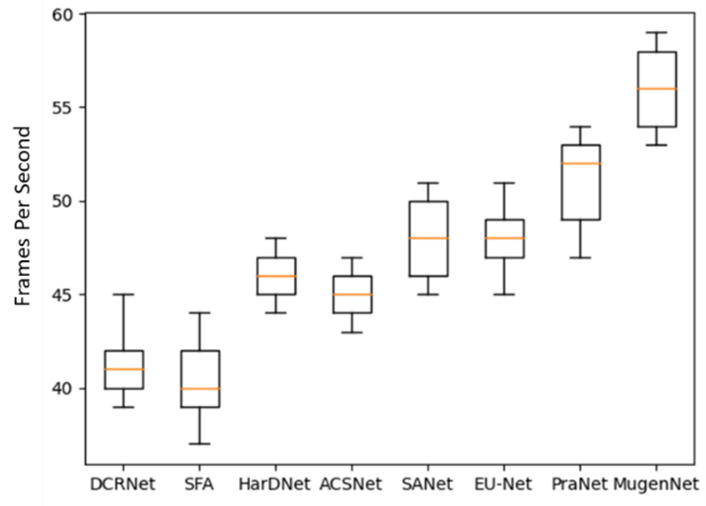}
    \caption{Training process on four datasets.}
    \label{fig fig.6}
\end{figure}

\subsection{Ablation Studies}
In this section, we conducted five sets of ablation experiments to evaluate the effectiveness of each component of MugenNet on two different datasets. The results are shown in Table \uppercase\expandafter{\romannumeral 7}, where TB represents the Transformer branch, CB represents the CNN branch, and MM represents the Mugen module.

\begin{table}[htbp]
\centering
\caption{COMPARISON OF THE RUNNING SPEED OF DIFFERENT MODELS}
\begin{tabular}{cccccc}
\hline
Methods           & Epoch & LR   & Time                 & FPS         & mDice \\ \hline
U-Net             & 30    & 3e-4 & $\sim$40 min         & $\sim$8fps  & 0.823 \\
U-Net++           & 30    & 3e-4 & $\sim$45 min         & $\sim$7fps  & 0.794 \\
SFA               & 500   & 1e-2 & \textgreater{}60 min & $\sim$40fps & 0.700 \\
PraNet            & 20    & 1e-4 & $\sim$30 min         & $\sim$50fps & 0.899 \\
\textbf{MugenNet} & 30    & 1e-4 & $\sim$15 min         & $\sim$56fps & 0.875 \\ \hline
\end{tabular}
\end{table}

\begin{table}[htbp]
\centering
\caption{ABLATION STUDIES OF MUGENNET}
\begin{tabular}{ccccccc}
\hline
\multirow{2}{*}{Settings} & \multicolumn{3}{c}{CVC-ColonDB}                                                                  & \multicolumn{3}{c}{ETIS-LaribPolypDB}                                                            \\ \cline{2-7} 
                          & mDice          & mIoU           & $F_\beta^\omega$ & mDice          & mIoU           & $F_\beta^\omega$ \\ \hline
Backbone                  & 0.519          & 0.473          & 0.637                                                          & 0.424          & 0.396          & 0.381                                                          \\
Backbone + TB             & 0.726          & 0.616          & 0.742                                                          & 0.649          & 0.605          & 0.578                                                          \\
Backbone + CB             & 0.667          & 0.591          & 0.680                                                          & 0.621          & 0.563          & 0.497                                                          \\
Complete Model            & \textbf{0.758} & \textbf{0.678} & \textbf{0.809}                                                 & \textbf{0.714} & \textbf{0.636} & \textbf{0.606}                                                 \\ \hline
\end{tabular}
\end{table}

We used three indicators to evaluate the models, i.e., $mDice$, $mIoU$ and $F_\beta^\omega$. Among them, $mDice$ and $mIoU$ were used to evaluate the accuracy of the models, while $F_\beta^\omega$ was used to evaluate the Precision and Recall of the models.

We respectively removed the Transformer branch, CNN branch and Mugen module from MugenNet to examine their performances. The result shows that our model (MugenNet) significantly improves the accuracy of semantic segmentation. Compared with the model without the Transformer branch, our model performs better on the ColonDB dataset than the other two cases. The result shows that $mIoU$ increased from 0.591 to 0.678, and $mDice$ increased from 0.667 to 0.758. 

The results in Table \uppercase\expandafter{\romannumeral 7} show that our model can segment colonic polyp image accurately. Our model outperforms the other two models in the ablation study, where either the CNN or the Transformer was omitted. Compared with the backbone neural network, the performance of our model in $mIoU$ improved about 43.34\% on the tested dataset (CVC-ColonDB). The results prove the superiority of our model (MugenNet).

\section{Conclusion}
In this paper, we proposed a new neural network (MugenNet), which combines Transformer and CNN. MugenNet shows excellent results in performing colonoscopy polyp segmentation tasks. We tested MugenNet on five polyp segmentation datasets (CVC-300, CVC-ClinicDB, CVC-ColonDB, ETIS-LaribPolypDB, Kvasir). Compared with traditional CNNs, our model achieves higher accuracy and faster processing speed. On the ETIS dataset, the average Dice can reach 0.714, which is 13.7\% higher than the best-performing model in current literature. Further, the experiments have shown that our model has a good generalization ability and has a robust performance on different datasets, compared with the other models (U-Net, U-Net++, ResUNet++, SFA, PraNet, DCRNet, EU-Net, SANet, ACSNet, HarDNet). Compared with traditional CNNs, our model (MugenNet) achieves faster training speeds. With the use of pre-trained weights from the DeiT distilled neural network, MugenNet converged in just 30 epochs. With a batch size of 16, it only takes 15 minutes, and the processing speed is about 56 fps. This performance in processing speed means that our model can be applied to video streaming data processing and real-time clinical polyp segmentation, laying the foundation for the next generation colonoscopy examination techniques.

In our experiment, we tested MugenNet on five public datasets (CVC-300, CVC-ClinicDB, CVC-ColonDB, ETIS-LaribPolypDB, Kvasir). The experiment results show that our model achieves good performance even when processing unknown image datasets. 

MugenNet is a neural network that combines CNN and Transformer. For the CNN branch, we chose ResNet-34, which can be replaced by other convolutional neural networks. We will test the impact of other CNN into MugenNet in the future. Further, we can optimize our model based on Model Parameters and Multiply-Accumulate Operations. We also aim to apply MugenNet to other applications such as lung infection diagnosis. 



\section*{Acknowledgement}
This work was supported in part by Natural Science Foundation of Shanghai through a Funding Grant (Grant No. 23ZR1416200).

\bibliographystyle{elsarticle-harv} 
\bibliography{reference}




%%\bibitem[()]{}

%%\end{thebibliography}
\end{document}